\documentclass{PoS}

\usepackage{multicol}

\let\OLDthebibliography\thebibliography
\renewcommand\thebibliography[1]{
  \OLDthebibliography{#1}
  \setlength{\parskip}{0pt}
  \setlength{\itemsep}{0pt plus 0.3ex}
}

\title{High resolution mm-VLBI imaging of Cygnus A}

\ShortTitle{High resolution mm-VLBI imaging of Cygnus A}

\author{\speaker{Bia Boccardi}$^a$%
         \thanks{Member of the International Max Planck Research School for Astronomy and Astrophysics at the Universities of Bonn and Cologne.}, Thomas Krichbaum$^a$, Uwe Bach$^a$, Eduardo Ros$^{abc}$, J. Anton Zensus$^a$\\ 
        $^a$: Max-Planck-Institut f\"{u}r Radioastronomie, Auf dem H\"{u}gel 69, D-53121 Bonn, Germany \\ 
        $^b$: Observatori Astron\`{o}mic, Universitat de Val\`{e}ncia, C.\ Catedr\'{a}tico Jos\'{e} Beltr\'{a}n 2, E-46980 Paterna, Val\`{e}ncia, Spain\\
        $^c$: Departament d'Astronomia i Astrof\'{\i}sica, Universitat de Val\`{e}ncia, C.\ Dr.\ Moliner 50, E-46100 Burjassot, Val\`{e}ncia, Spain\\
        E-mail: \email{bboccardi@mpifr-bonn.mpg.de}}

\abstract {At a distance of 249 Mpc ($z$=0.056), Cygnus A is the only powerful FR II radio galaxy for which a detailed sub-parsec scale imaging of the base of both jet and counter-jet can be obtained. Observing with VLBI at millimeter wavelengths is fundamental for this object, as it uncovers those regions which appear self-absorbed or free-free absorbed by a circumnuclear torus at longer wavelengths. 
We performed 7\,mm Global VLBI observations, achieving ultra-high resolution imaging on scales down to 90 $\mu$as. This resolution corresponds to a linear scale of only $\sim$400 Schwarzschild radii. We studied the transverse structure of the jets through a pixel-based analysis, and kinematic properties of the main emission features by modeling the interferometric visibilities with two-dimensional Gaussian components. Both jets appear limb-brightened, and their opening angles are relatively large ($\phi_\mathrm {j}\sim 10^{\circ}$). The flow is observed to accelerate within the inner-jet up to scales of $\sim$ 1 pc, while lower speeds and uniform motions are measured further downstream. A single component seen in the counter-jet appears to be stationary. These observational properties are explained assuming the existence of transverse gradient of the bulk Lorentz factor across the jet, consisting of a fast central spine surrounded by a slower boundary layer.}

\FullConference{12th European VLBI Network Symposium and Users Meeting,\\
		7-10 October 2014\\
		Cagliari, Italy}

\begin{document}
\section{Introduction}
The powerful radio galaxy Cygnus A ($z$=0.056, $D_{\mathrm{L}}$=249 Mpc) is a unique target for investigating the physical processes taking place at the very base of a relativistic jet. According to current magneto-hydrodynamic models and simulations \cite{Meier}, the crucial mechanisms for jet launching and acceleration happen on very small scales, of the order of tens to hundreds Schwarzschild radii, which can be reached by present-day mm VLBI arrays when observing the nearest sources. Global VLBI observations\footnote{\scriptsize{The European VLBI Network is a joint facility of European, Chinese, South African and other radio astronomy institutes funded by their national research councils. The National Radio Astronomy Observatory is a facility of the National Science Foundation operated under cooperative agreement by Associated Universities, Inc.}} of Cygnus A at 7\,mm (43 GHz) enable high resolution imaging on scales down to $\sim$90 $\mu$as, translating into a linear size of only 400 Schwarzschild radii 
at the source redshift (for $M_{\mathrm{BH}}=2.5 \times 10^9 M_{\odot}$ \cite{Thad}). The combination of the small observing beam and the high radio frequency provides a sharp view towards the central engine and unveils those regions which appear either self-absorbed or that are affected by free-free absorption from a molecular torus at lower frequencies (e.g. \cite{Krici}). Showing a prominent counter-jet also on parsec scale, Cygnus A is likewise ideal for testing the unified scheme of AGN \cite{Bart}, with the additional advantage that both the jet and the counter-jet can be transversally resolved. While spatial resolution is essential for shedding light on the aforementioned topics, such detailed imaging pays back by revealing a very complex picture of the internal physics occurring in the plasma flow. In particular, there is increasing evidence that jets are not homogeneous outflows, but can show a pronounced transverse stratification. For example, limb brightened jets have been observed on pc-scale 
both in some nearby FR I radiogalaxies, like M\,87 \cite{Yuri} and 3C\,84 \cite{Nagai} and in blazars, as in the case of Mrk 501 \cite{Giro} and 3C\,273 \cite{Andrei}. This feature has been interpreted with the existence of a spine-sheath structure in the jet, comprising a central, light and ultra-relativistic spine surrounded by a layer of slower and denser material \cite{Sol}.  Whether this structure arises directly from the jet formation process or is instead the result of instabilities in the flow, is still unclear. Anyhow, it is important to keep in mind that, because of relativistic effects, the presence of a transverse gradient of the bulk Lorentz factor can strongly affect the observational properties of jets. This is especially true for sources like Cygnus A whose jets are seen at a large viewing angle. In this 
case, the fast part of the flow can be completely invisible, both in jet and counter-jet, because of relativistic de-boosting (Fig. 1). In the following we present a study of the jet transverse structure and kinematics in Cygnus A from Global VLBI observations at 7\,mm, showing that Doppler de-boosting may indeed play a major role.
\begin{figure}[!h]
\centering
\includegraphics[trim=0.5cm 0.5cm 0.1cm 0.5cm, clip=true, scale=0.1]{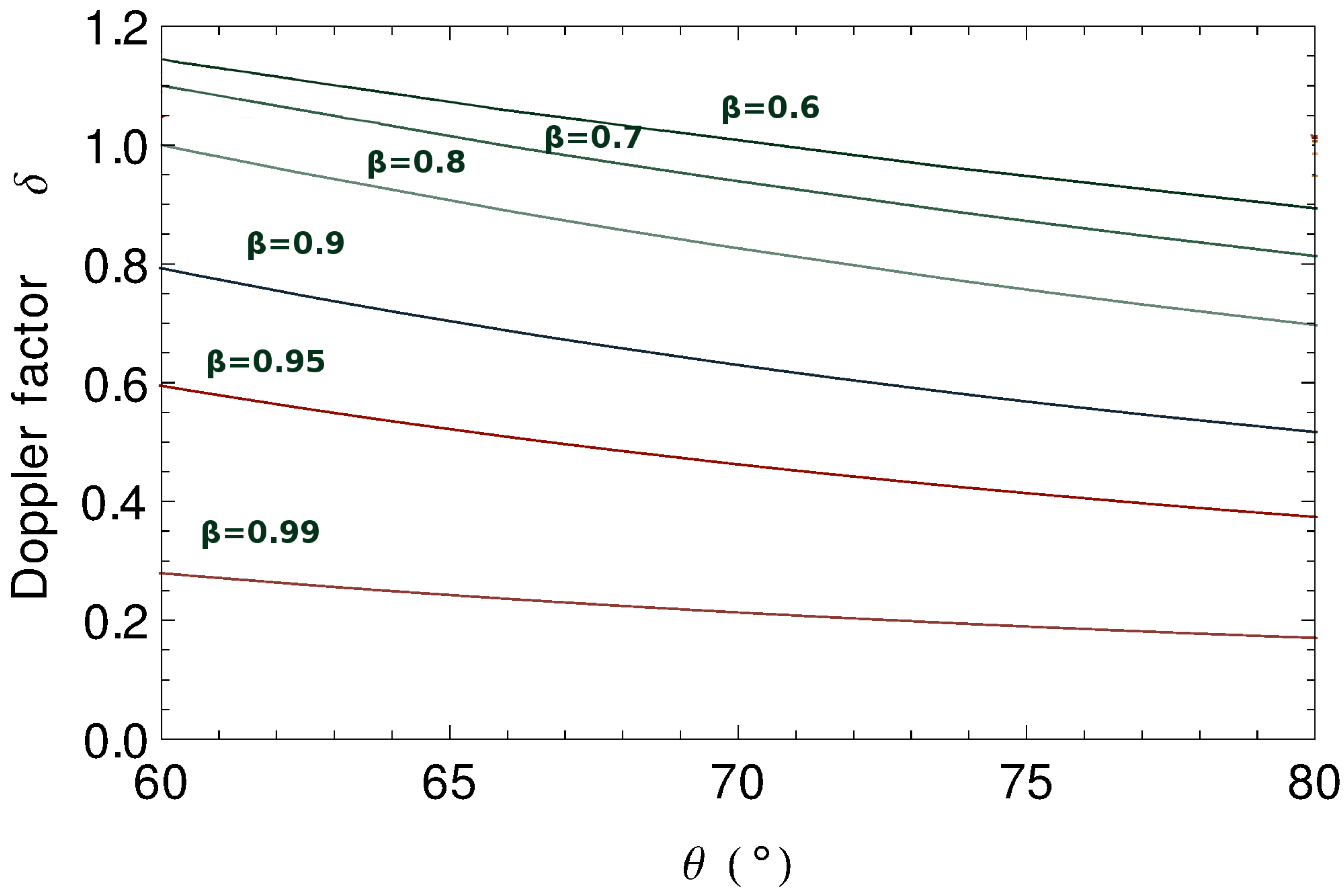}\\
\vspace{-0.15cm}
\caption{\footnotesize{Doppler factor $\delta$ versus (large) viewing angle $\theta$ for different intrinsic speeds $\beta$.}}
\end{figure}
\section{Data set}
Our Cygnus A data set comprises four epochs from Global VLBI observations made between 2007 and 2009 (see Table 1). For the first three epochs, data were recorded in dual polarization mode using 4 subbands (IFs) with a total bandwidth of 64 MHz per polarization (512 Mb/s recording rate). The last epoch was observed in single polarization, with 16 subbands and a total bandwidth of 128 MHz. The data were calibrated in \textsc{AIPS} following the standard procedures, while imaging and self-calibration of amplitude and phase were performed in \textsc{DIFMAP}. In Fig. 2, we present an image obtained by stacking the \textsc{CLEAN} images from the four epochs, after restoring them with a common circular beam of 0.1 mas. The source shows at this frequency a complex structure, with an extended counter-jet and a prominent limb-brightening of the jet on mas scales.
\begin{table}[!t]
\caption{\footnotesize{Log of observations and map characteristics. Col. 1: Day of observation. Col. 2: VLBA - Very Long Baseline Array; GBT - Green Bank Telescope; On - Onsala; Nt - Noto; Eb - Effelsberg; Yb - Yebes. Col. 3: Beam and position angle. Col. 4: Peak flux density. Col. 5: Map noise. All values are for untapered data with uniform weighting.}}
\centering
\footnotesize
\begin{tabular}{c c c c c }
\hline
\hline
Date       & Antennas  & Beam FWHM $\mathrm{[mas, deg]}$& S$_{\mathrm {peak}} \mathrm {[mJy/beam]}$&rms$\mathrm {[mJy/beam]}$                \\
\hline
23/10/2007 & VLBA, GBT, Yb, Eb, On, Nt &$0.23\times 0.11, -21.9^{\circ}$&176&0.13 \\
16/10/2008 & VLBA, GBT, Yb, Eb, On, Nt &$0.22\times 0.10, -11.0^{\circ}$&256&0.11 \\ 
19/03/2009 & VLBA, GBT, Yb, Eb, On, Nt &$0.23\times 0.09, -12.5^{\circ}$&264&0.14 \\
11/11/2009 & VLBA, GBT, Yb, Eb, On, Nt &$0.19\times 0.10, -20.1^{\circ}$&235&0.32 \\ 
\hline
\end{tabular}
\end{table}
\begin{figure}[!h]
\centering
\includegraphics[width=\textwidth]{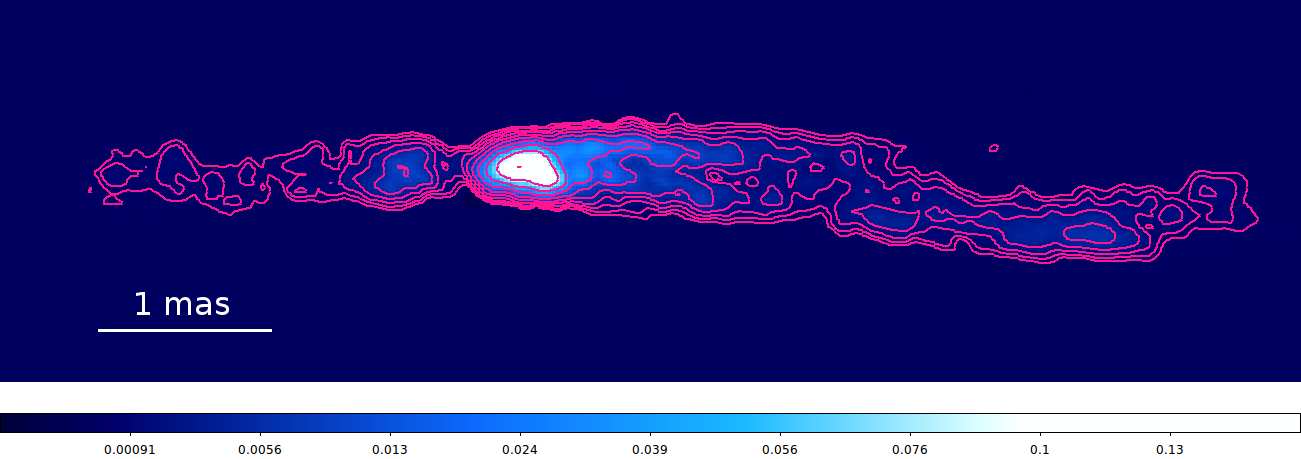}\\
\hspace{14cm}\tiny{\textbf{[Jy]}}
\caption{\footnotesize{Stacked 43 GHz image from observations in 2007-2009, rotated by 16$^{\circ}$ clockwise. It is convolved with a circular beam of 0.1 mas FWHM. Contours represent isophotes at 0.3, 0.6, 1.2, 2.4, 4.8, 9.6, 19.2, 38.4, 76.8, 153.6 mJy/beam. The four maps were aligned as described in Fig. 5.}}
\end{figure}
\section{Ridge line}
We carried out a detailed analysis of the jets transverse structure and opening angle for a single epoch (November 2009). 
The image was restored with a circular beam of 0.15 mas FWHM, rotated by 16$^{\circ}$ and sliced pixel-by-pixel (1 px=0.03 mas) in the direction perpendicular to the jet axis. 
Then, a Gaussian fit of the brightness distribution along the $y$ axis was performed twice in each slice, first fitting only one Gaussian and then fitting two Gaussian profiles. We confirm earlier findings \cite{Uwe} of a double ridge line on the jet side, and we also find evidence for it in the counter-jet (Fig. 3). Here the limb brightening is less pronounced than in the jet side, the two rigde lines are closer, but still clearly separated after accounting for positional errors, estimated to be 1/5 of the FWHM of the respective component. In the outer-jet the northern rail appears to fade, while the southern rail is visible up to a core separation $r \simeq -$3.5 mas.
\begin{figure}
\centering
\includegraphics[trim=0.5cm 1cm 2cm 2cm, clip=true, scale=0.27]{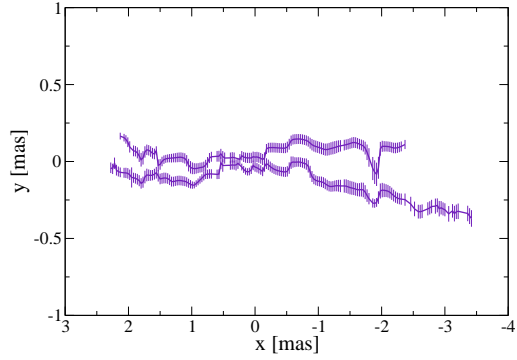}
\caption{\footnotesize{Double ridge line in jet and counter-jet in the 43 GHz map from November 2009. Vertical lines are errorbars on the peak positions of the fitted Gaussians, and are set to 1/5 of the FWHM of the respective component. Note that the $x$ and $y$ scales are different.}}
\end{figure} 
\section{Opening angle}
The jet width was calculated for each slice as the separation between the edges of the two (deconvolved) FWHM. The results are shown in Fig. 4 (left), where the deconvolved widths obtained from a single-Gaussian slice fit are also plotted for comparison, and show good agreement. Both the jet and the counter-jet appear conical, with some minor oscillations of the width, up to $r\simeq \pm$ 1.7 mas, where a prominent narrowing takes place. For $|r|<1.7$ mas, the half opening angle $\phi^{\mathrm {app}}$ was calculated by performing a linear fit to the jet width (Fig. 4, right), which shows for the counter-jet a factor of two narrower opening angle than for the jet:
\begin{equation}
\phi^{\mathrm {app}}_{\mathrm {j}}=\left(5.10 \pm 0.16\right)^{\circ}, ~~~~~~
\phi^{\mathrm {app}}_{\mathrm {cj}}=\left(2.42 \pm 0.21\right)^{\circ}
\end{equation}
\begin{figure}[!h]
\centering
\includegraphics[trim=0cm 0cm 0cm 2cm, clip=true, scale=0.26]{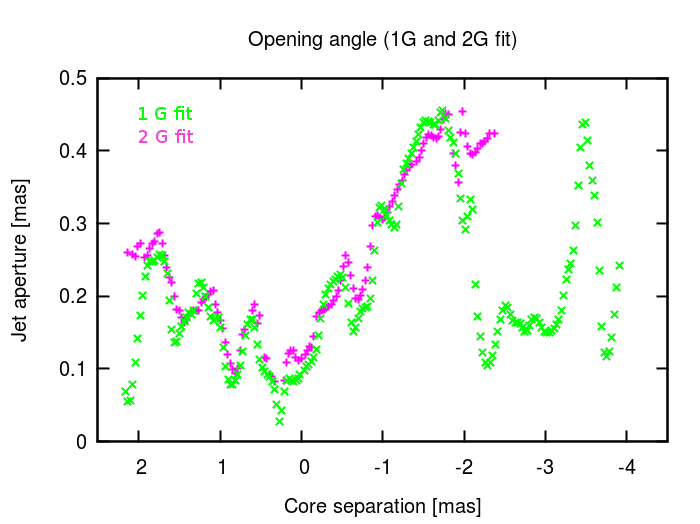} \quad
\includegraphics[trim=0cm 0cm 0cm 2cm, clip=true, scale=0.26]{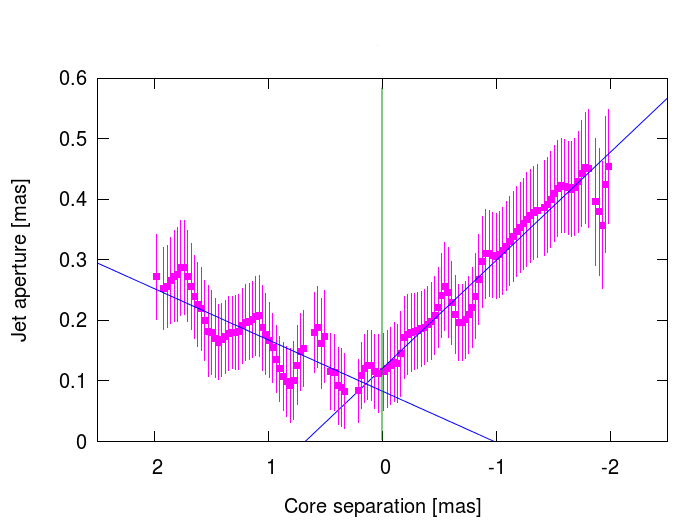}
\caption{\footnotesize{\textbf{Left -} Jet width versus core separation as deduced from fitting two Gaussian functions (magenta) or a single Gaussian (green). \textbf{Right -} Half apparent opening angles calculated from a linear fit to the jet width.}}
\end{figure} 
In the classical scenario of intrinsic symmetry of the jet and the counter-jet, their apparent opening angle should, in principle, be the same, unless they are strongly misaligned. A small misalignment is indeed present in Cygnus A both on pc and kpc scales, but it is not enough to justify the observed difference. An alternative explanation may invoke the difference between the signal-to-noise ratios (SNR) of approaching and receding sides: the considerably dimmer counter-jet may indeed appear to be narrower than the brighter jet. However, the hypothesis of intrinsic asymmetry cannot be readily discarded with the present data. As it was also found in other radio galaxies, the opening angles in Cygnus A are considerably larger than what it is found in blazars, typically featuring full intrinsic opening angles of $(1-2)^{\circ}$ \cite{Push}. Assuming a reasonably large viewing angle of $70^{\circ}$, the intrinsic full opening angle of Cygnus A is $\sim10^{\circ}$ on the jet side. This result supports the spine-
sheath scenario, in which sources seen at larger viewing angles become more and more sheath-dominated, and therefore broader, while the emission from blazars mostly comes from the strongly boosted, thin spine. 
\section{Kinematic structure}
Finally, we present a study of the kinematic properties of the radio emission in the jets during the 2-year monitoring interval. In order to obtain a simplified model of the source, describing only the main emission features, we have performed a \textsc{MODELFIT} analysis in \textsc{DIFMAP}, i.e we fit circular Gaussian patterns to the self-calibrated visibilities (Fig. 5, left). The images were aligned to the position of the component N, assumed to be stationary. The same choice was made in \cite{Krici}, also based on compactness and spectral index arguments. According to our cross-identification, a new component (J1) was ejected between October 2007 and March 2009, giving rise to a brightening of about +50\% of the core and inner-jet region.
\begin{figure}[!h]
\centering
\includegraphics[trim=0cm 4.8cm 7.5cm 0cm, clip=true, scale=0.16]{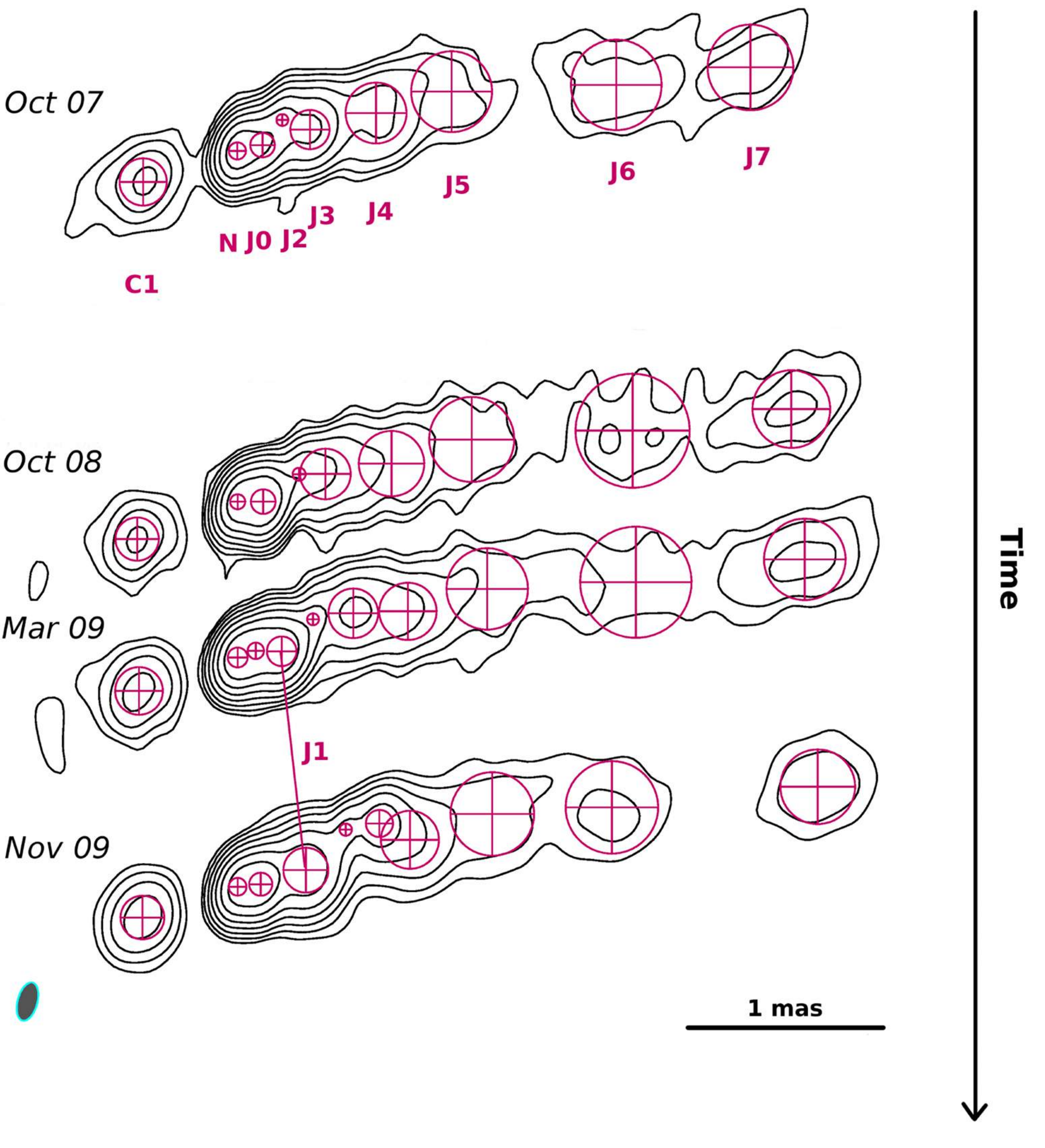} \quad
\includegraphics[trim=0cm 0cm 0cm 0cm, clip=true, scale=0.085]{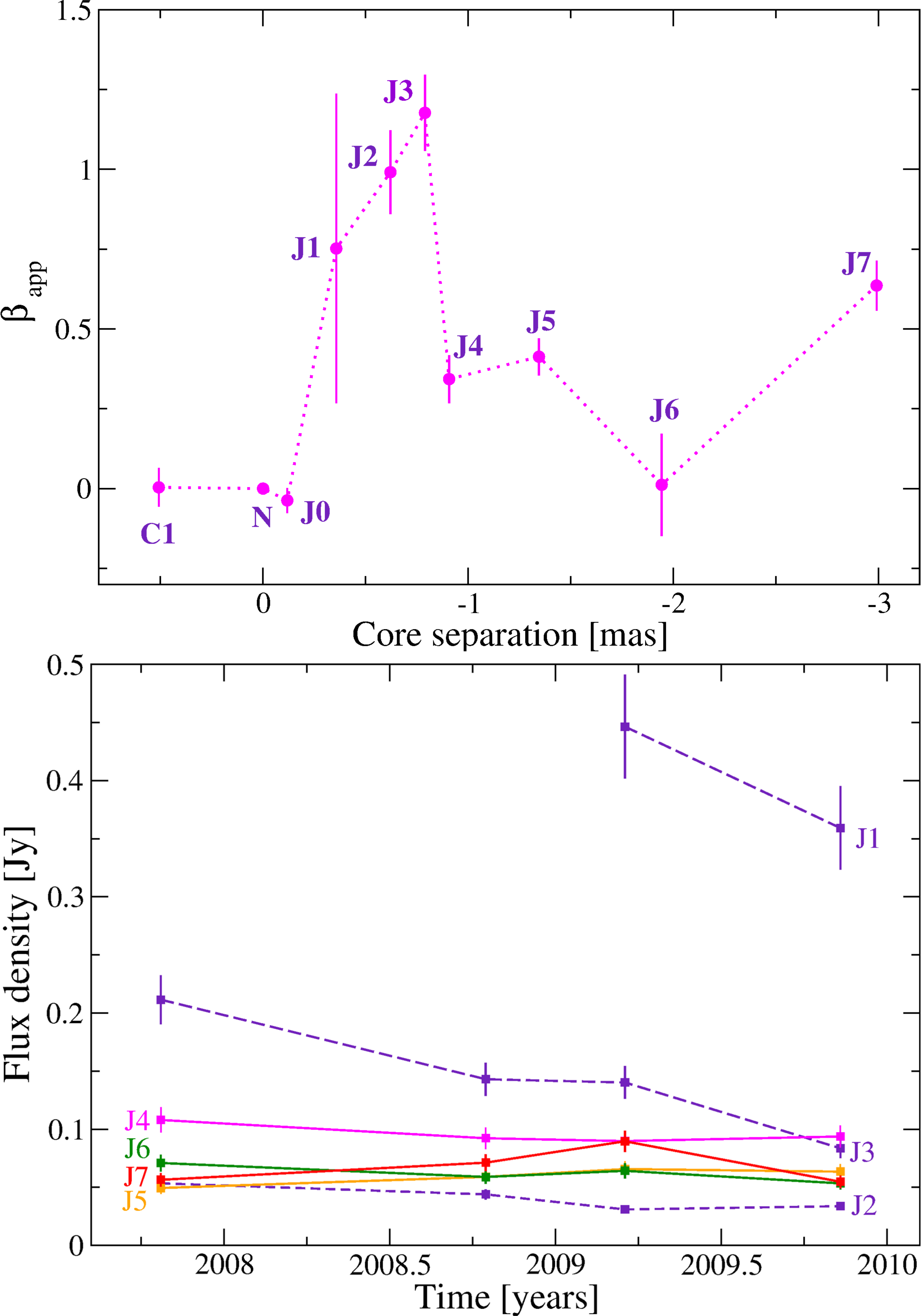}
\caption{\footnotesize{\textbf{Left -} 43 GHz modelfit maps of Cygnus A. Images were tapered and convolved with a common beam FWHM of 0.2 $\times$ 0.1 mas in P.A. -15$^{\circ}$. Contours represent isophote curves at 1.8, 3.6, 7.2, 14.4, 28.8, 57.6 and 115.2 mJy/beam. The reference point for the alignment is component N, assumed to be stationary. \textbf{Top right -} Apparent speed versus core separation. \textbf{Bottom right -} Light curves of the modelfit components. }}
\end{figure}
The proper motion $\mu$ of components J2 and J3 is best described by quadratic functions, indicating an acceleration, while features in the outer-jet move each at a roughly constant speed. The apparent speed $\beta_{\mathrm {app}}= v_{\mathrm{app}}/c$ can be calculated as $\beta_{\mathrm {app}} = (\mu D_L)/(c(1+z))$, where $D_L$ the luminosity distance, $c$ is the speed of light and $z$ is the redshift.  
The dependence of $\beta_{\mathrm {app}}$ on the core separation is shown in Fig. 5 (top right). In the approaching jet we observe an acceleration up to $\beta_{\mathrm {app}} = 1.18 \pm 0.12$ within $\sim$ 0.8 mas from the core. Then the speed drops significantly, remaining largely sub-luminal in the outer-jet. At a core separation $r\simeq -2$ mas, we observe a stationary feature that approximately coincides with the position of the narrowing in Fig. 4 (left). Component C1 in the counter-jet does not show a significant proper motion ($\beta_{\mathrm {app}} = 0.004 \pm 0.060$). These characteristics can be well explained by the stratification scenario. By looking at the light-curves of the components in Fig. 5 (bottom right), we notice that those crossing the acceleration region (J1, J2, J3) also show a decreasing flux density with time, while flux densities are more constant for components in the outer-jet. This could be another piece of evidence for the existence of a fast spine, which is 
accelerating and 
therefore getting fainter, until the emission gets fully dominated by the slower layers of the sheath. This also naturally explains the observed limb-brightening. A similar argument can be applied to the counter-jet: only the very slow, and therefore the less de-boosted part of the flow can be seen.
\section{Conclusions}
 We have presented preliminary results from a high resolution millimeter VLBI study of the radio galaxy Cygnus A. We have shown that the observed limb brightening and the kinematic properties of the jet, as well as its large opening angles, can be well explained in a scenario of transverse jet stratification. The impact of stratification on the observational properties of jets depends on the viewing angle, and it is certainly strong for radiogalaxies. 

\end{document}